\documentclass[b5paper,twoside]{jpconf_pre}
\usepackage{graphicx}

\begin{document}
\title[GRB follow-up in Korea]{Gamma-Ray Burst Afterglow Follow-up Observation in Korea}

\author[M.Im]{Myungshin Im}

\address{CEOU/Astronomy Program, Dept. of Physics \& Astronomy, Seoul National University, Seoul, KOREA}

\ead{mim@astro.snu.ac.kr}

\begin{abstract}
We review Gamma-Ray Burst (GRB) afterglow follow-up observations being carried out by our group in Korea. We have been performing GRB follow-up observations using the 4-m UKIRT in Hawaii, the 2.1-m telescope at the McDonald observatory in Texas,  the 1.5-m telescope at Maidanak observatory in Uzbekistan, the 1.8-m telescope Mt. Bohyun Optical Astronomy Observatory (BOAO) in Korea, and the 1.0-m remotely operated telescope in Mt. Lemmon, Arizona. We outline our facilities, and present highlights of our work, including the studies of high redshift GRBs at $z > 5$, and several other interesting bursts. 
\end{abstract}

\section{Introduction}

GRBs are the most energetic explosions in the universe which result from deaths of stars in special ways. Their enormous energy output ($E_{iso} \sim 10^{53} \, erg\, s^{-1}$) produces very bright afterglows in UV/optical/NIR which can be followed up using moderate-sized ground-based telescopes to understand the nature and the origin of GRB and the emission mechanisms of the afterglow. The bright afterglows serve as a useful tool to understand the distant universe since they could be observed beyond $z > 10$, deep into the re-ionization epoch. They offer valuable information on the re-ionization state and the dust origin in the early universe, as well as the metal enrichment history over the cosmic time. 
In Korea, we started a GRB follow-up observation from 2007 October \cite{Lee10}. Initially, we focused on the study of the early universe using GRBs as bright probes, but the program evolved over the time, expanding the facilities and the scientific scope of our studies. Here, we describe our GRB follow-up observation activities.  

\section{Facilities}

Figure 1 and Table 1 summarize the facilities used by our group for the GRB afterglow observation. The facilities are scattered around the world, allowing round-the-clock observation of GRB afterglows. Emphasis is given to NIR imaging to study high redshift GRBs, with the imagers equipped with NIR filters (at least $z$ and $Y$). Three facilities are the observational facilities of our group at CEOU/SNU. We have regular access to the UKIRT and its wide-field NIR imager, WFCAM, for about 20-50 nights per year. We installed SNUCAM \cite{Im10}, an optical imager, on the 1.5m telescope of the Maidanak observatory, Uzbekistan, and it has been used for GRB observations since 2006. Recently, we started using an optical CCD camera, CQUEAN (Camera for QUasars in the EArly Universe) for the 2.1m telescope in McDonald observatory \cite{Park11,Kim11}. In addition, we are regular users of the 1.8m telescope at BOAO, 0.6m telescope at SOAO, and 1.0m telescope at LOAO, all of which are operated by the Korea Astronomy \& Space Science Institute (KASI). 

\vspace{-0.3cm}

\begin{center}
\begin{table}[h]
\caption{Our GRB follow-up facities}
%\footnotesize\rm
\centering
\begin{tabular}{llll}
\br
Facility & Instrument & Filters & Location \\
\mr
UKIRT 4m & WFCAM & ZYJHK & Hawaii, US \\
McDonald 2.1m & CQUEAN & grizY,is,iz & Texas, US \\
Maidanak 1.5m & SNUCAM (4k$\times$4k) & UBVRI,ugrizY & Uzbekistan \\
BOAO 1.8m & KASINICS/CCD cam. & JHK/UBVRI & Mt. Bohyun, Korea \\
LOAO 1.0m & 4k$\times$4k CCD cam. & UBVRIzY & Mt. Lemmon, US \\
SOAO 0.6m & 2k$\times$2k CCD cam. & UBVRI & Mt. Sobek, Korea \\
\br
\end{tabular}
\end{table}
\end{center}

\vspace{-0.5cm}

\begin{figure}[h]
\begin{center}
\includegraphics[width=6.5cm,angle=270]{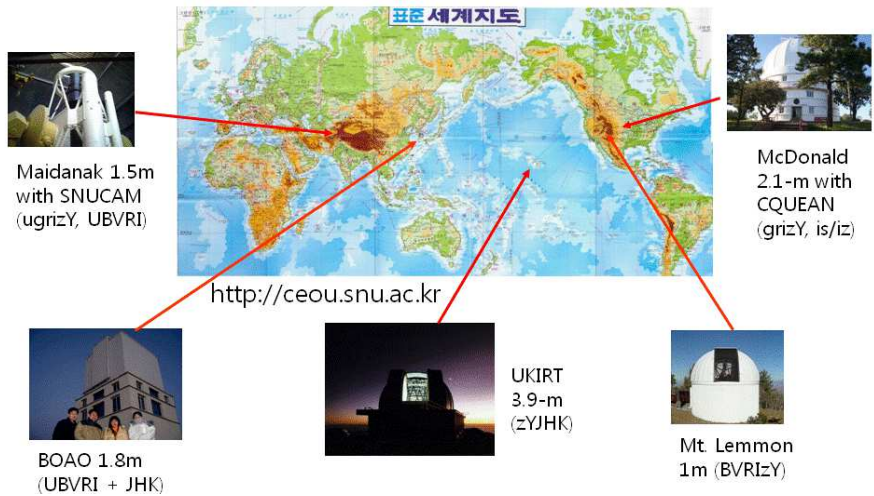}
\caption{\label{label} Our GRB follow-up facilities. All of the facilities here are capable of NIR imaging at least out to $Y$-band.}
\end{center}
\end{figure}

\vspace{-0.3cm}

\section{GRBs of special interest}

 The GRB follow-up activities have been ongoing since the observation of our first target, GRB 071010B \cite{Lee10,Urata09}. We have been observing about 10 GRBs per semester, and most of the observations have been reported in the GCN Circulars. The GRB observation activity using the LOAO 1-m telescope is described in detail in \cite{Lee10}, and our activity follows this framework. We describe below a few GRB studies to highlight our activity.

\subsection{GRB 071025}

 This GRB was observed on 2011 October 25/26 using LOAO 1-m telescope. We took $BVRI$ images starting from 20 minutes after the BAT alert. It is found that GRB 071025 is a dust-reddened GRB at $z \sim 5$ \cite{Jang11,Perley10}. We were the first group to report its high redshift nature, based on a red color of $R - I > 2$ mag \cite{Im07,Lee10} (Figure 2). Analysis of its spectral energy distribution (SED) suggests that the dust originates from a core-collapse supernova. Currently, GRB 071025 stands as the most convincing case among high redshift GRBs to harbor SN-dust \cite{Jang11,Perley10}. 

\begin{figure}[h]
\begin{minipage}{5.4cm}
\includegraphics[width=5.4cm,angle=0]{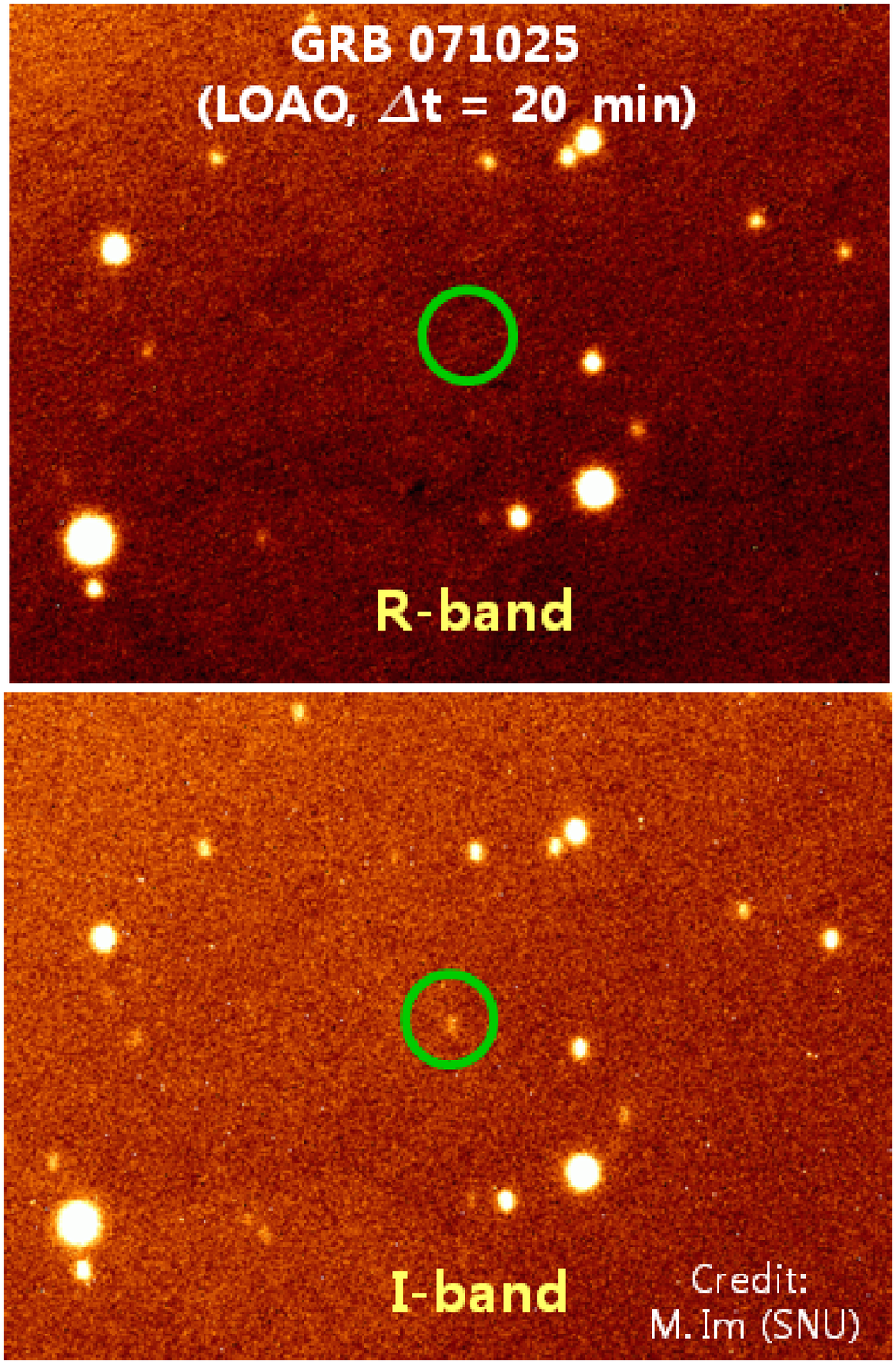}
\caption{\label{label} The LOAO images of GRB 071025 in $R$ and $I$. The clear detection in $I$, but not in $R$, suggests a break in SED due to its high redshift ($z=5$).}
\end{minipage}\hspace{0.5cm}%
\begin{minipage}{8.0cm}
\includegraphics[width=8.0cm]{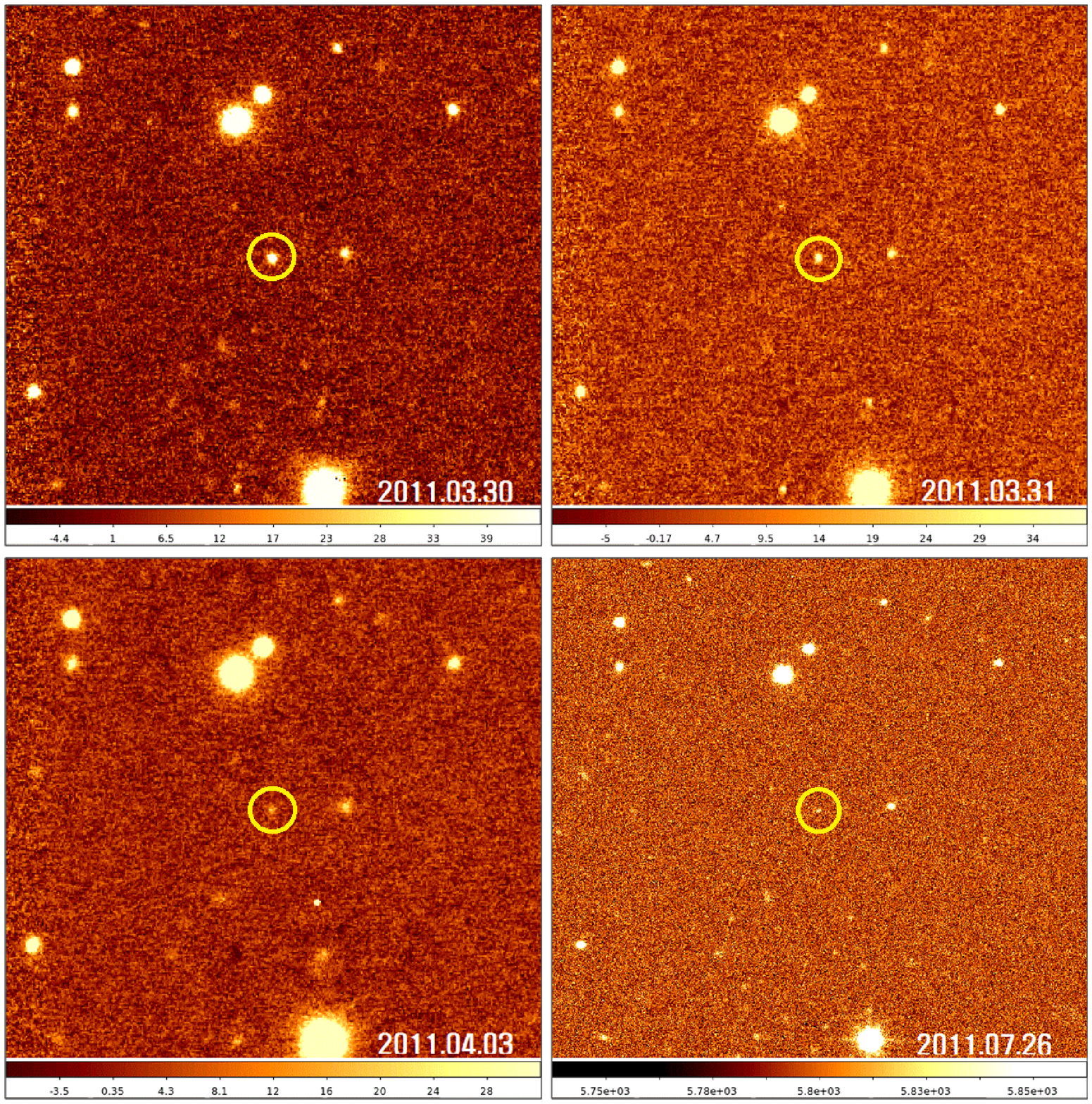}
\caption{\label{label} The K-band images of Swift J1644+57 taken by the BOAO 1.8-m and UKIRT (bottom-right) telescopes. Swift J1644+57 is an object at the center of the circle, and the numbers in the bottom right indicate the dates of the observation.}
\end{minipage}\hspace{-1cm}
\end{figure}

\subsection{GRB 080319B}

 GRB 080319B was a very bright GRB, with its peak magnitude reaching about $V = 6$ mag (named as "naked eye" GRB). We obtained optical imaging data at LOAO and Maidanak, which were used to constrain the physical parameters and the emission mechanism of the burst \cite{Pandey09}. 

\subsection{GRB 090423A}

 GRB 090423A is currently the most distant GRB with spectroscopic redshift at $z=8.2$ \cite{Tanvir09,Salvattera09}. We followed up this object in $IzY$ using LOAO and Maidanak facilities, and derived $Y$-band field calibration data which were useful for unveiling the high redshift nature of the burst \cite{Im09,Tanvir09}.  

\subsection{GRB 100905A}

 GRB 100905A was observed at UKIRT in $zJHK$, 15 minutes after the BAT alert \cite{Im10b}.
 Our analysis of the NIR photometry data reveals a sharp break between $z$-band and $J$-band data,
 suggesting that this is a very high redshift GRB at $z=7.5 \pm 0.7$.

\subsection{GRB 101225A}

 This burst appeared on the Christmas night of 2010. It had an unusually very long burst duration, and exotic behavior in the X-ray and the optical light curves. We observed the burst in $rizY$ using CQUEAN on the McDonald 2.1m telescope, 0.3 days after the BAT alert or shortly after the observers finishing their Christmas dinner. A comprehensive analysis including our data suggests that this is a new kind of GRB whose afterglow is dominated by a black-body, created in a He star-neutron star merger \cite{Thone11}.  

\subsection{Swift J164449.3+573451 (Swift J1644+57)}

 Swift J1644+57 was discovered on 2011 March 28 as a GRB with a very long burst duration. Soon, it was realized that this is not a GRB, but a very interesting phenomenon occurring at the center of a galaxy at $z=0.354$. We performed extensive follow-up observations in NIR and optical using all of our follow-up facilities. A multi-wavelength analysis of the data led to a conclusion that the onset of Swift J1644+57 was caused by a jet-emission from a tidal disruption event where a star came too close to a supermassive black hole with $M \sim 10^{7}\, M_{\odot}$ at the center of the host galaxy \cite{Burrows11}.

\ack

 This work was supported by the Korea Science and Engineering Foundation (KOSEF) grant No. 2010-0000712, funded by the Korean government (MEST). I thank the collaborators of our follow-up observations, especially, W.-K. Park, Y. Jeon, C. Choi, M. Jang, I. Lee, Y. Urata, K. Huang, M. Ibrahimov, J. Youn, I. Baek, S. Pak, H. Jeong, J. Lim, Y.-B. Jeon, and H.-I. Sung for their help with the analysis of the data and the observations at various facilities.


\section*{References}
\begin{thebibliography}{12} 
\bibitem{Lee10} Lee, I., Im, M., \& Urata, Y. 2010, Journal of the Korean Astronomical Society, 43, 95
\bibitem{Im10} Im, M., et al. 2010, Journal of the Korean Astronomical Society, 43, 75
\bibitem{Jeon10} Jeon, Y., Im, M., Ibrahimov, M., et al. 2010, ApJS, 190, 166
\bibitem{Park11} Park, W.-K., Pak, S., Im, M., et al. 2012, PASP, in press
\bibitem{Kim11} Kim, E., Park, W.-K., et al. 2011, Journal of the Korean Astronomical Society,  44, 115
\bibitem{Urata09} Urata, Y., K. Huang, Im, M., et al. 2011, ApJ, 706, L183
\bibitem{Jang11} Jang, M., Im, M., Lee, I., et al. 2011, ApJ, 741, L20
\bibitem{Perley10} Perley, D., Bloom, J. S., Klein, C. R., et al. 2010, MNRAS, 406, 2473
\bibitem{Im07} Im, M., Lee, I., \& Urata, Y. 2007, GCN Circ. 6994, 1
\bibitem{Pandey09} Pandey, S. B., Castro-Tirado, A. J., Jelinek, M., et al. 2009, A\&A, 504, 45
\bibitem{Tanvir09} Tanvir, N. et al. 2009, Nature, 461, 1254
\bibitem{Salvattera09} Salvaterra, R., et al. 2009, Nature, 461, 1258
\bibitem{Im09} Im, M., Jeon, Y., Choi, C., et al. 2009, GCN Circ. 9242
\bibitem{Im10b} Im, M., Choi, C., Jun, H., et al. 2010, GCN Circ., 11222
\bibitem{Thone11} Thone, C. et al. 2011, Nature, 480, 72
\bibitem{Burrows11} Burrows, D. et al. 2011, Nature, 476, 421
\end{thebibliography}
\end{document}